\documentclass[a4,preprint]{article}

\usepackage{amssymb}
\usepackage{amsmath}
\usepackage{latexsym}

\usepackage[switch]{lineno}

\usepackage{url}
\usepackage{xcolor}
\usepackage{graphicx}
\definecolor{newcolor}{rgb}{.8,.349,.1}

\usepackage[citebordercolor=white]{hyperref}

\usepackage{algorithm}
\usepackage{algpseudocode}

\begin{document}

\title{Physically-Informed Fuzzy Clustering of Vertical Sounding Ionograms} 

\author{Oleg I.Berngardt,  Sergey N.Ponomarchuk}

\maketitle

\begin{abstract}
This paper presents a physically-informed fuzzy clustering of
vertical sounding ionograms for automatically separating the ionogram
into tracks suitable for further interpretation and determining their
optimal number. The model is designed for use not only in conditions
where the number of tracks is known, but also in disturbed ionospheric
conditions where the number of tracks is preliminary unknown.

The method is based on an expectation-maximization algorithm, used for
clustering, and on parametrically specified distributions
of distances from points to parametrically specified curves. The curves
used as track models are close to model tracks in the
parabolic ionospheric layer model. The resulting model of
each track has six parameters: three standard ones (the critical frequency,
the lower boundary of the layer, and its half-width), 
and three additional ones to take into account possible
underlying layer effects. By sequentially increasing the number of
tracks and optimizing their parameters, the model finds the optimal
number of tracks on the ionogram by minimizing the modified Bayesian information
criterion. The Sequential Least Squares Quadratic Programming algorithm is used
to find the parameters of a single track. The width
of each single track is assumed to be unknown constant found during
fitting process.

To improve the quality of ionogram clustering, automatic adaptive
noise filtering is performed before clustering. This filtering is
based on a combination of the DBSCAN and Gaussian Mixture algorithms.
Also, to improve clustering quality on an ionosonde without hardware separation
of the ordinary and extraordinary components, a preliminary approximate
removal of points belonging to the extraordinary mode is performed.
\end{abstract}


\section{Introduction}

Automatic processing of vertical sounding ionograms is one of important problems 
in ground-based ionospheric diagnostics. A vertical
sounding ionogram represents the signal amplitude $A(f,r)$ as a function
of sounding frequency f and effective range r, corresponding to the
group delay of the reflected signal. Usually, ionograms contain
several tracks - the points with signifficant amplitude, grouped near
a curve describing dependence of effective range on frequency. The
number and the shape of the tracks are determined by radio wave propagation
conditions: the shape of the ionospheric layer, which depends on external
conditions and processes in the ionosphere; the geographic location
and antenna pattern of the ionosonde; the orientation of the magnetic
field during radio wave propagation in an anisotropic medium;
the absorption level in the lower ionosphere; and ambient noise, which
limit the number, visibility, and intensity of observed tracks due
to the finite sensitivity of the receiver and finite power of transmitter.

\begin{figure}
\centering
\includegraphics[scale=0.65]{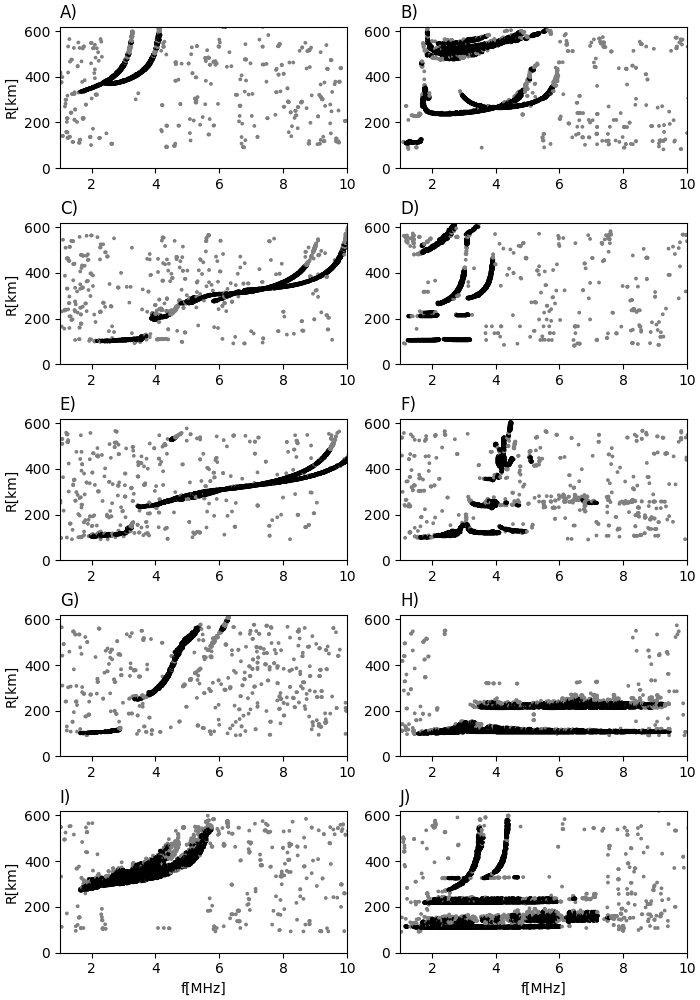}
\caption{Examples of ionograms obtained with the ISTP SB RAS vertical sounding
ionosonde. On the left (A, C, E, G) are ionograms consistent with
the model of a one-dimensionally inhomogeneous ionosphere in the absence
of semi-transparent layers and multiple reflections; on the right
(B, D, F, H, J) the ionograms that do not fit this model. Points presumably
containing noise are highlighted in gray, while informational points
are highlighted in black (noise selection was performed using the
automatic algorithm described below in this paper).}
\label{fig:1}
\end{figure}

Early studies of automatic digital processing of ionograms used relatively
simple computational algorithms: a low-parameter quasi-parabolic model
of the F2 layer and the corresponding model track \cite{Reinisch_1983},
or an equivalent greedy search for merging discontinuous tracks into
single one using a sector of possible directions continuing the curve \cite{Fox_1989}.
These methods proved effective in determining the shape and parameters of simple
ionograms, primarily F-layer tracks. Later, more sophisticated approaches
were developed
to fit the shape of the ionogram F-track \cite{PEZZOPANE2004,Ding_2007,PEZZOPANE2008,ZHENG2013,Chen_2013,JIANG2015,Song_2016,JIANG2017968}
by parametric functions.
An another approach is fast extracting
from the ionogram a small number of parameters of the F2
layer, which are important for forecasting
the  radio wave propagation conditions, for example foF2 (or MUF)
\cite{IPPOLITO2015}, or foF2, hmF2 \cite{LYNN2018}.

However, the real ionosphere is vertically and horizontally irregular
and is not described by single F layer. Therefore, with the growth
of computing power, solutions for processing the ionograms
not descibed in terms of simple models have been made: for example,
the models for identifying the sporadic E-layer (Es) \cite{Scotto_2007},
or for separating F, spread-F, and Z-tracks \cite{Scotto_Pezzopane_2012}.
More complex modern models can calculate the parameters of a given
number of layers - Es, E, F1, F2 \cite{Heitmann2019} or E,F1,F2 \cite{Ponomarchuk2026}. 
One of the
most modern approaches in this direction is training neural networks
of various architectures on pre-labeled datasets to classify ionogram points into:
E and F-layers \cite{Kvammen_2024}, E, F1, F2 layers \cite{XIAO2020},
or E, F, Es, spread-F tracks \cite{Rao_2022}. Usually, multiple
reflections are not taken into account when processing ionograms by most models.

Most of the papers above are based on the assumption that the ionosphere
is described by a parametric model specified by the model developers:
the electron density depends only on altitude, semi-transparent Es
may exist, but complex semi-transparent layers or multiple reflections,
from Es or F layer, are absent. Such simplified models can be reffered as a 1.5D
inhomogeneous ionosphere. In this case, the model ionogram (excluding
semi-transparent layers and spread-F) can be considered as a mapping
$f\rightarrow r_{o},r_{x}$, i.e., each frequency corresponds to no
more than two reflection heights - for the ordinary and extraordinary
components of the radio wave (the extraordinary component sometimes
absent). After eliminating semi-transparent Es traces, the problem
of reconstructing the profile parameters $\theta$ can be solved in
a simple problem statement, for example, as minimizing the residual
functional between observations and the model:

\begin{equation}
\Omega=\sum_{f_{i},r_{i}\in O}W(f_{i},r_{i})\left(r_{o,i}^{(model)}(f_{i}|\theta)-r_{i}\right)^{2}+\sum_{f_{i},r_{i}\in X}W(f_{i},r_{i})\left(r_{x,i}^{(model)}(f_{i}|\theta)-r_{i}\right)^{2}\rightarrow min
\label{eq:SimpleFit}
\end{equation}

where O, X are sets of points identified as O and X components of
the signal, and W - some weights.

In this paper we propose a solution the problem in different, wider
formulation. In a disturbed 3D-inhomogeneous ionosphere, the number
of tracks observed at a fixed frequency can be arbitrary. Additional
tracks can arise due to reflections from semi-transparent and sporadic
layers of varying multiplicity (e.g., at ranges of 100, 200, and 300
km for Es); due to multiple reflections from F-layers at ranges of 500-600
km; due to presence
of horizontal inhomogeneities, forming several signal propagation
trajectories for each mode. Disturbances of the three-dimensional
ionosphere can also arise, and produce additional tracks. 
Tracks of different modes can contact and/or
intersect, and therefore a simple mapping $f\rightarrow r_{o},r_{x}$ (or $R\rightarrow R^{2}$)
describing the ionogram does not exist. Accordingly, a simple approach
like (\ref{eq:SimpleFit}) will not work.

Fig.\ref{fig:1} shows examples of vertical sounding ionograms obtained
on the ISTP SB RAS vertical ionosonde at the mid-latitude Tory observatory
(52N,103E) \cite{Kurkin2024}, and used further in the paper to illustrate
the proposed technique.
On the left in Fig.\ref{fig:1} are shown ionograms that correspond
the 1D-inhomogeneous ionosphere model in the absence of complex
layers and multiple reflections (\textasciitilde 30\% for the studied
dataset), on the right - those that do not fit such a model (\textasciitilde 70\%
for the studied dataset). The gray color indicates presumably noise
points, the black color indicates information points; noise selection
was performed by the automatic algorithm described below in this paper.
On the right are shown (from top to bottom): 
complex double reflections from the F-layer in the presence of inhomogeneities; 
double reflections from the F-layer and from the semi-transparent sporadic E-layer; 
multiple reflections from the E- and Es-layers; 
total absorption of signals in the semi-transparent Es-layer with double reflections;
multiple reflections from semi-transparent Es-layer, reflection from E-layer inhomogeneities, and reflection from F-layer.

Therefore, in a disturbed ionosphere, instead of the well-known problem
of finding the parameters of a fixed model, a more complex problem
arises: the combined detection of each track shape, estimating parameters
of each track, and finding the number of tracks that is statistically
optimal for ionogram description as a superposition
of certain physically interpretable tracks. 
If we consider the ionogram
as points in a three-dimensional space of sounding
frequencies, effective ranges, and signal amplitudes $(f_{i},r_{i},A(f_{i},r_{i}))$,
then the problem can be formulated as fuzzy model clustering
of these points, with the possibility of intersection of different
clusters, their contact, and varying point densities within a cluster.
The shape of each cluster must correspond to a certain
physical model of a track so that an expert can later
interpret it. So, the problem belongs to
the class of problems of physically-informed fuzzy clustering.

From the experimenter's perspective, the program must determine from
the ionogram how many physically interpretable tracks it contains
and associate each ionogram point with the probability that it belongs
to each particular track. The interpretation of each track
will be a separate task, beyond the scope of this paper. Therefore,
the aim of our paper is to develop an algorithm for optimal
recognizing an unknown number of tracks in a vertical sounding ionogram.

\section{Method description}

\subsection{Noise filtration}

Vertical sounding ionosondes are sensitive transmit-receiving systems
that receive signals both from itself and from external sources. External
signals can be separated into regular (concentrated interference from
radio broadcasting stations) and irregular (noise-like) ones. One approach
to remove the external signals is to divide the ionograms into blocks (f,r) and remove
the noise from each block individually, followed by morphological fitting
of a continuous curve to the model track. This is effectively suitable for
determining the parameters of narrow, simple ionograms, primarily
with F-layer track \cite{Chen_2013}.

Below we assume that concentrated interference from
ionograms has been already removed using, for example,
\cite{Podlesnyi_2014} algorithm, and low-intensity noise signals
has been also removed by threshold processing. Therefore, we further assume
that the data contains only ionospheric signals (ionogram tracks)
and random noise. The noise on ionograms appears as isolated dots,
while tracks appear as dense clusters of dots. Therefore, it looks
useful to use density-based methods to detect and remove the noise as outliers in
the data.

One such widely known method is Density-Based Spatial Clustering of Applications with Noise (DBSCAN) method  
\cite{DBSCAN_1996,DBSCAN2_2017},
which allows to separate the data into clusters based on their degree
of isolation (distance to the nearest neighbor) and suitable for
arbitrary cluster (i.e.track) shapes. 
This method has two hyperparameters: the maximum distance for combining two
closest points into a single cluster ($\varepsilon$) and the minimum number
of points in a cluster required to identify the cluster as separate
one ($n_{min}$).

\subsubsection{Minimum number of points in a cluster $n_{min}$.}

Points combined into very small clusters ($n<n_{min}$) by the DBSCAN algorithm
are considered outliers in the same way as completely isolated points.
The hyperparameter $n_{min}$ determines the minimum number of points
in a cluster that the algorithm can distinguish from noise. Obviously,
the more complex the cluster model (the more internal parameters it
has), the more points are needed to determine these parameters, and
the higher $n_{min}$ should be. If the shape of the cluster (track)
is determined by K parameters, then $n_{min}>=K$. Below we use 6-parameter
tracks, therefore $n_{min}=10$ is chosen. This allows us to interpret
clusters with fewer than 10 points as outliers, and clusters with
at least 10 points as tracks or their parts. This method imposes a
lower limit on the amount of data that allows interpreting an ionogram
track: very short tracks of less than 10 points will be ignored at
the adaptive filtering stage as outliers. Another disadvantage of
the method is that it marks the edges of tracks as noise, cutting
off the corresponding number of edge points.

\subsubsection{Minimum distance between cluster points $\varepsilon_{opt}$.}

The ionogram can thus be separated into points grouped into clusters
larger than $n_{min}$ points each, and noise-like outliers. The maximum
possible distance to the nearest neighbor in a cluster is determined
by the hyperparameter $\varepsilon$. Unfortunately, our ionogram
is a rectangle $[f_{min},f_{max}],[r_{min},r_{max}]$, and the scales
and units of measurement along the different axes are different -
MHz for the x-axis, and kilometers for the y-axis. Therefore,
the distance calculations are arbitrary. Since ionograms are traditionally
interpreted visually using images with a 1:1 aspect ratio, to better
match the expert assessment, the distance was assumed to be Euclidean,
and a preliminary scaling of the ionogram to zero mean and unit variance
was performed for all available ionogram points for each coordinate
independently:

\begin{equation}
\begin{array}{c}
\tilde{f}=\frac{f-\overline{f}}{\sigma(f)}\\
\tilde{r}=\frac{r-\overline{r}}{\sigma(r)}
\end{array}\label{eq:norm_prefilter}
\end{equation}

where $\overline{f},\sigma(f)$ are the sample mean and standard deviation
of the sounding frequency for the ionogram points, similarly for $\overline{r},\sigma(r)$.
In the normal distribution approximation, the scaled effective ranges
and scaled sounding frequencies are in the range $\tilde{f}\in[-3,3],\tilde{r}\in[-3,3]$.
It should be noted that we use the scaling only when identifying
outliers (noise); at further stages of ionogram processing, the original
sounding frequencies and effective ranges are used.

The data consists of two classes: nearby points that
form tracks, and isolated points that represent noise. The cutoff
distance $\varepsilon$ is defined as the cutoff distance that optimally
separates these two classes. Typically, when solving clustering problems,
the hyperparameter $\varepsilon$ is selected either manually or based
on some clustering quality metric, such as the Silhouette coefficient,
the elbow method, etc.

We will use different approach. By analogy with the Local Outlier Factor (LOF)
method \cite{LOF_2000} and to speed up the calculations, we assume
that the probability density $P(\rho_{i})$ of the average Euclidean
distances $\rho_{i}$ between points in a two-dimensional space $d_{ij}$,
taken over the 10 ($n_{min}$) nearest neighbors $N_{10}(i)$ for each point $i$
is:

\begin{equation}
\begin{array}{c}
d_{ij}=\left\Vert \left(\tilde{f}_{i},\tilde{r}_{i}\right),\left(\tilde{f}_{j},\tilde{r}_{j}\right)\right\Vert _{2}=\sqrt{\left(\tilde{f}_{i}-\tilde{f}_{j}\right)^{2}+\left(\tilde{r}_{i}-\tilde{r}_{j}\right)^{2}}\\
\rho_{i}=\frac{1}{10}\sum_{j\in N_{10}(i)}d_{ij}
\end{array}\label{eq:dist_prefilter}
\end{equation}

and has the form of a weighted sum of two normal distributions $\mathcal{N}(\rho|\mu_{j},\sigma_{j})$ 
- one for signals, another for noise:

\begin{equation}
\begin{array}{c}
\rho_{i}\sim P(\rho)=\eta_{0}\mathcal{N}(\rho|\mu_{0},\sigma_{0})+\eta_{1}\mathcal{N}(\rho|\mu_{1},\sigma_{1})\\
\eta_{0}+\eta_{1}=1
\end{array}\label{eq:model_prefilter}
\end{equation}

where the parameters of these normal distributions $\mu_{0},\sigma_{0},\mu_{1},\sigma_{1}$
and their weights $\eta_{0},\eta_{1}$ are determined by fitting the
experimental distance distribution $\rho_{i}$ (that is, to find these
parameters, we use clustering by a mixture of Gaussian distributions,
GM\cite{GM_1996}). The optimal threshold $\varepsilon_{opt}$ is
defined as the optimal boundary of membership in each of these Gaussian
distributions, satisfying the equation:

\begin{equation}
\eta_{0}\mathcal{N}(\varepsilon_{opt}|\mu_{0},\sigma_{0})=\eta_{1}\mathcal{N}(\varepsilon_{opt}|\mu_{1},\sigma_{1})\label{eq:eps_prefilter}
\end{equation}

\begin{figure}
\centering
\includegraphics[scale=0.6]{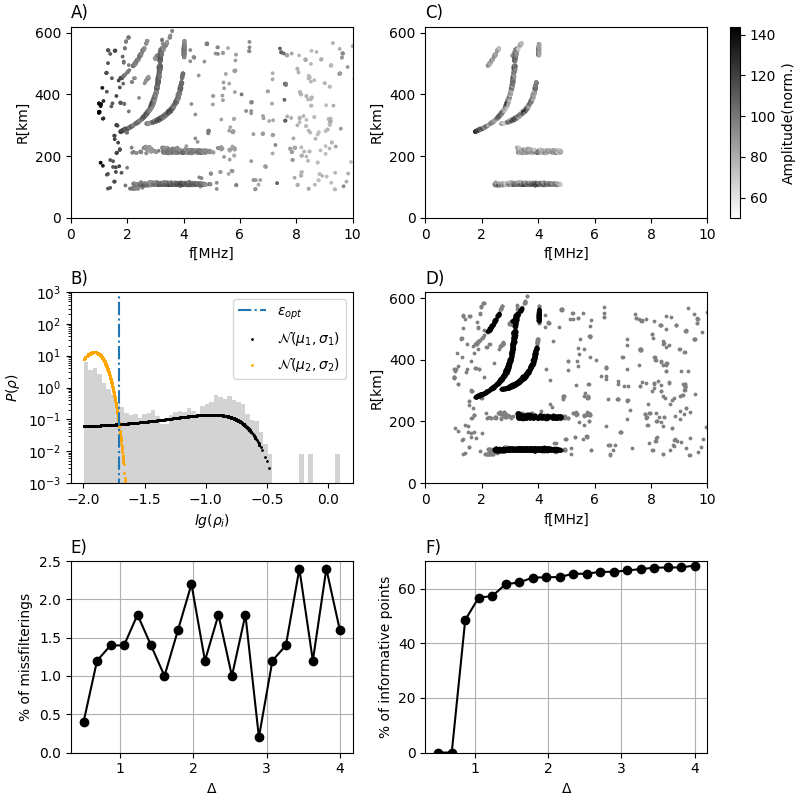}
\caption{Filtering algorithm. A) the original thresholded ionogram; B) the
distribution of distances $\rho_{i}$ for it, the result of its approximation
by the model and the position of the optimal boundary $\varepsilon_{opt}$;
C) the resulting filtered ionogram; D)
Labeling of noise (gray) and information (black) points 
in accordance with the $DBSCAN(\varepsilon_{opt},n_{max})$
algorithm; E) dependence of the percentage of noise
misfiltering on $\Delta$ ; F) the percentage of points on the ionogram, 
detected as informational ones.}
\label{fig:filtering} 
\label{fig:delta_dependence}
\end{figure}

The resulting hyperparameters $\varepsilon_{opt},n_{max}$ are used
for the DBSCAN algorithm: points not included into clusters are considered
noise, while those included in clusters are considered informational ones.
Fig.\ref{fig:filtering}A-D shows the algorithm's output:
the original ionogram after thresholding, 
the resulting ionogram with noise removed,
the mean distance to 10 nearest neighbours distribution
$\rho_{i}$, 
the result of labeling into noise and information signals by the suggested algorithm.

We can estimate how often the filtering fails when the threshold level
changes by scaling the DBSCAN hyperparameter $\varepsilon$ in $\Delta$ times:

\begin{equation}
\varepsilon=\Delta\cdot\varepsilon_{opt}\label{eq:delta_dependence}
\end{equation}

The choice of the threshold value $\varepsilon_{opt}$ is stochastic
due to the stochastic nature of the GM algorithm implementation (usually
it is based on the EM algorithm \cite{Dempster_1977}),
and therefore may occasionally make errors. The dependence of the
final percentage of noise misfiltering cases on $\Delta$ is shown
in Fig.\ref{fig:filtering}E. The percentage of 
points on the ionogram detected as informative ones 
is shown in Fig.\ref{fig:filtering}F. It is
evident from the figure that the optimal value of $\Delta$ , on the
one hand, should not fail very often and on the other hand should select a
sufficiently large number of informative points on the ionogram. So $\Delta$ should be
in the range {[}1,1.7{]}. Therefore, our choice of $\varepsilon_{opt}$
(or $\Delta=1$) as the hyperparameter is acceptable and causes the
misfiltering error rate of about $10^{-2}$.

To compensate misfiltering due to the stochastic
nature of the GM algorithm, the threshold calculation and subsequent
clustering using the DBSCAN method are performed twice, and the variant
that detects more isolated points (noise) is selected. This significantly
reduces the probability of misfiltering to approximately $10^{-4}$. 
Instead of DBSCAN method one can use LOF method with found $\varepsilon_{opt}$ 
hyperparameter - the results will be close.

\subsection{Single track model}

To separate tracks, it is necessary to have a model for the shape
of each track and a method for finding its parameters. Most tracks
can be divided into ordinary and extraordinary wave tracks,
having close shapes. 
Let's consider the simplest track model
- vertical reflection of ordinary radiowave from a parabolic ionospheric layer with the
following dependence of the squared plasma frequency $f_{model}^{2}(h)$ (proportional to electron density)
on altitude h:

\begin{equation}
f_{model}^{2}(h)=\left\{ \begin{array}{l}
f_{0}^{2}\left(1-\frac{(h-h_{0})^{2}}{y_{m}^{2}}\right);h\in[h_{0}-y_{m},h_{0}+y_{m}]\\
0;h\notin[h_{0}-y_{m},h_{0}+y_{m}]
\end{array}\right.\label{eq:qp_layer_model}
\end{equation}

The parameters of this parabolic layer are $f_{0}$ - the maximum
plasma frequency of the layer, $h_{0}$ - the height of the layer
maximum, and $y_{m}$ - the layer half-thickness.

The ionogram track $R(f)$ is a curve that
approximately describes the mean positions of the ionogram point heights
at a fixed sounding frequency,
and corresponds the integral:

\begin{equation}
\begin{array}{c}
R(f)=\int_{0}^{h_{r}}\frac{1}{\sqrt{1-f_{model}^{2}(h)/f^{2}}}dh\\
h_{r}=min_{h}(f_{model}^{2}(h)=f^{2})
\end{array}\label{eq:qp_integral}
\end{equation}

where $h_{r}$ is the reflection height of the radio wave.

For a parabolic layer, the ionogram track has the 
known analytical form\cite{Davies1966,Reinisch_1983}:

\begin{equation}
\begin{array}{l}
R_{O}(f)=(h_{0}-y_{m})+y_{m}\cdot f_{1}\cdot atanh(f_{1})\\
f_{1}=\frac{f}{f_{0}}\in[0,1)
\end{array}
\label{eq:qp_track_model_o}
\end{equation}

The model has three parameters: $h_{0}$, $y_{m}$, and $f_{0}$. For
the extraordinary component, in a first approximation, for the same
electron density profile shape, we can get track shape by replacing
plasma frequency $f_{0}\leftarrow f_{0}+f_{H}$, where $f_{H}$ is
the frequency dependent on the geomagnetic field and the ionosonde's
latitude. Therefore, both tracks (the ordinary and extraordinary
waves) can be described by the same model with different parameters.

This model has two problems: the profile shape near the electron
density peak is not described by a parabola, and the model is unsuitable
for multiple layers. To improve the model, the following track model
was chosen:

\begin{equation}
\begin{array}{l}
R_{model}(f|\theta)=h_{1}+y_{m}\cdot f_{1}\cdot atanh(f_{1})+A\cdot y_{m}\cdot(B-f_{1})\cdot atanh(B-f_{1})\\
f_{1}=\left(\frac{f}{f_{0}}\right)^{C},\theta=(h_{1},y_{m},f_{0},A,B,C)
\end{array}\label{eq:track_model}
\end{equation}

\begin{figure}
\centering
\includegraphics[scale=0.6]{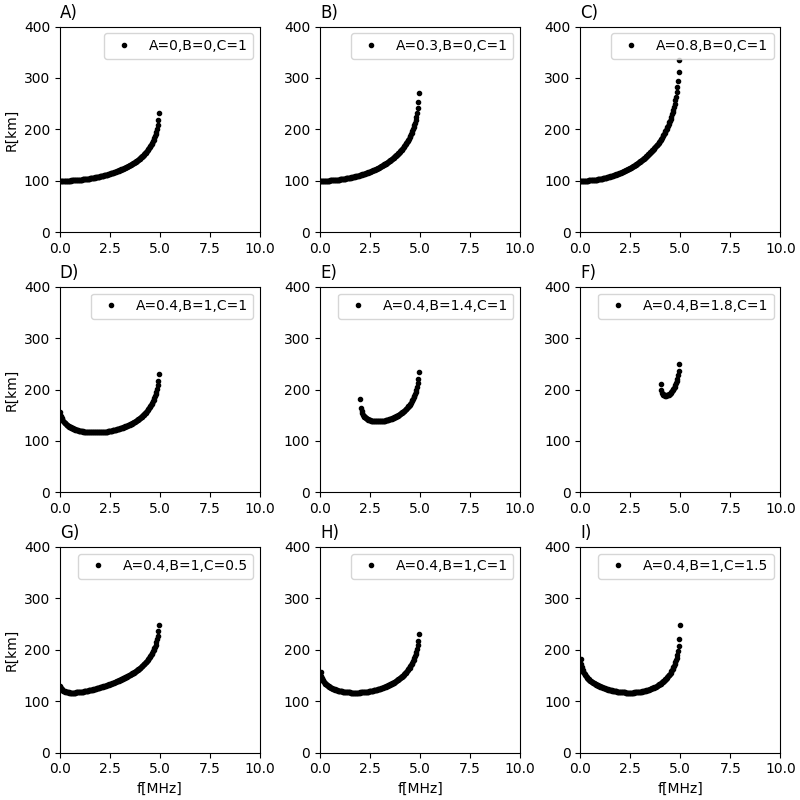}
\caption{Examples of tracks with different values of parameters A,B,C. The
remaining parameters are fixed:$h_{1}=100,y_{m}=50,f_{0}=5$}
\label{fig:track_shape} 
\end{figure}

Here, to simplify the model, the parameter $h_{1}=h_{0}-y_{m}$ is
used. The third term in the model is related to the possible presence
of an underlying layer - in our analysis, we attempt to analyze reflection
from different layers separately. The resulting model has six parameters
and is flexible enough to describe many observed tracks. Examples
of tracks with different parameter values and the corresponding ionograms
are shown in Fig.\ref{fig:track_shape}.

From general physical considerations and statistical characteristics
of the ionosphere, the model parameters vary: $h_{1}$ within the
range of $[min(R)/2,max(R)+1]$km, $y_{m}$ within $[0..500]$ km,
and $f_{0}$ within $[1..25]$MHz. From Fig.\ref{fig:track_shape}A-C
it is evident that parameter A has the meaning of the ratio of half-thicknesses
in layers, it is responsible for the rate of change of the track with
frequency and varies within $[0..1]$. From Fig.\ref{fig:track_shape}D-F
it is evident that parameter B determines the position of the low-frequency
part of the track, is responsible for the ratio of plasma frequencies
and heights with the unfitted underlying layer and varies within $[0..5]$.
Fig.\ref{fig:track_shape}G-I shows that the parameter C determines
the shape of the low-frequency and partially high-frequency parts
of the track, has the meaning of the difference of electron density
shape from the parabola in the region of the layer maximum, and varies within
the range {[}1, 1.5{]}. This model describes two of the four basic
track shapes proposed in \cite{Chen_2018}.

\subsection{Single track parameters search}

For a real-time algorithm, the use of an accurate and fast inversion
algorithm is critical. We fit the experimental track $(f,r,A(f,r))$
to the $R_{model}(f|\theta)$ model and determine the model parameters $\theta$
defined by the 6-dimensional vector $\theta=[h_{1},y_{m},f_{0},A,B,C]$.
Minimization of the loss function was used for the inversion. For
an initial comparison of the clustering quality, the following loss
functions were tested:

\begin{equation}
\begin{array}{c}
WMAE(\theta)=L_{1}(\theta)=\frac{\sum_{i=1}^{N}A_{i}\left|r_{i}-R_{model}(f_{i}|\theta)\right|}{\sum_{i=1}^{N}A_{i}}\rightarrow min\\
WMSE(\theta)=L_{2}(\theta)=\frac{\sum_{i=1}^{N}A_{i}\left|r_{i}-R_{model}(f_{i}|\theta)\right|^{2}}{\sum_{i=1}^{N}A_{i}}\rightarrow min\\
WMAPE(\theta)=L_{3}(\theta)=\frac{\sum_{i=1}^{N}A_{i}\left|\frac{r_{i}-R_{model}(f_{i}|\theta)}{r_{i}}\right|}{\sum_{i=1}^{N}A_{i}}\rightarrow min\\
L_{4}(\theta)=\frac{\sum_{i=1}^{N}A_{i}\left|log(r_{i})-log(R_{model}(f_{i}|\theta))\right|}{\sum_{i=1}^{N}A_{i}}\rightarrow min\\
L_{5}(\theta)=\frac{\sum_{i=1}^{N}A_{i}\left|\frac{r_{i}-R_{model}(f_{i}|\theta)}{\sqrt{r_{i}}}\right|}{\sum_{i=1}^{N}A_{i}}\rightarrow min\\
L_{6}(\theta)=\frac{\sum_{i=1}^{N}A_{i}\left|r_{i}-R_{model}(f_{i}|\theta)\right|^{3/2}}{\sum_{i=1}^{N}A_{i}}\rightarrow min
\end{array}\label{eq:track_model_fit_loss}
\end{equation}

Tracks can be approximated by minimization of various losses with a weight corresponding to the
amplitude of each ionogram point $A_{i}=A(f_{i},r_{i})$. It should
be noted that the problem (\ref{eq:track_model_fit_loss}) must be
solved with the constraints associated with the physicality of the
model parameters discussed above.

An analysis of the final clusterer's results showed that the best
clustering quality of the analyzed losses (\ref{eq:track_model_fit_loss})
was achieved when using the WMAE (weighted MAE) loss functional. 
The other loss functionals from (\ref{eq:track_model_fit_loss})
produce worse results.

\begin{figure}
\centering
\includegraphics[scale=0.5]{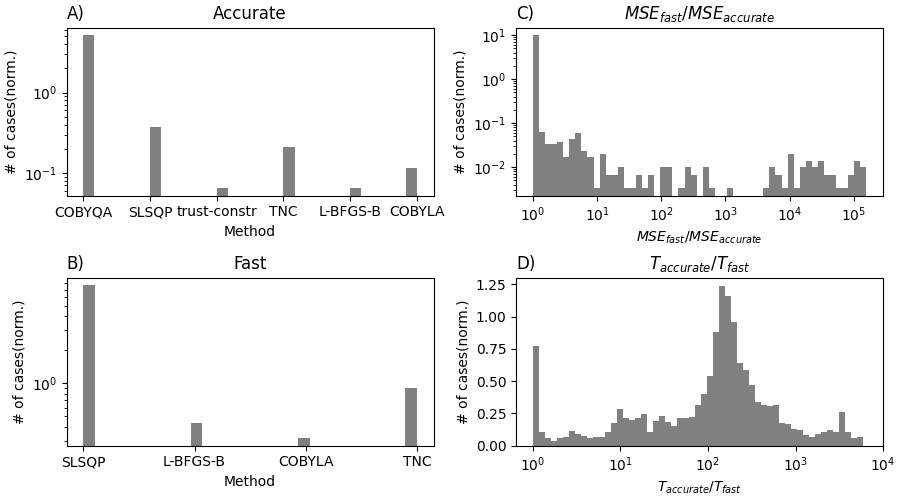}
\caption{Comparison of various numerical methods for searching the model track
parameters in terms of their speed and accuracy for real tracks inversion.
A) Distribution of cases when a method is found to be the most accurate;
B) Distribution of cases when a method is found to be the fastest;
C) Distribution of MSE relations between fastest and most accurate methods;
D) Distribution of execution time relations between  most accurate and fastest methods.
}
\label{fig:cmp_inv_algo} 
\end{figure}

To select the optimal algorithm for searching track parameters, a
comparison of 6 algorithms for minimizing the nonlinear functional
$L(\theta)$ with constraints was carried out on several thousand
real tracks: Sequential Least Squares Quadratic Programming (SLSQP) \cite{Lawson_and_Hanson_1995},
Trust region method with constraints (Trust-constr) \cite{Conn_et_al_2000}, Truncated
Newton Conjugate-Gradient method (TNC) \cite{NASH200045}, Limited-memory
Broyden--Fletcher--Goldfarb--Shanno with Bound constraints method
(L-BFGS-B) \cite{Byrd_1996}, Constrained Optimization BY Linear Approximation
(COBYLA) \cite{Powell1994}, Constrained Optimization BY Quadratic
Approximations (COBYQA) \cite{COBYQA}. For each experimental track,
the fastest and the most accurate (in terms of MSE) search algorithm were determined,
and the frequency distribution of different models among the fastest
and most accurate was plotted. Additionally, for each track the two
ratios have been claculated: between the accuracies (i.e. MSE) of the accurate
and fast algorithms, and between the fitting times of the fast and accurate
algorithms. Fig.\ref{fig:cmp_inv_algo} shows the comparison results.
The figure shows that COBYQA is the most accurate model, with SLSQP
second in accuracy. SLSQP is the fastest model, TNC is the second
one. It should be noted that the ratio of forecast errors between
the fast and accurate models is quite close to 1, while the ratio
of execution time between the accurate and fast models is of the order
100. Therefore, to solve the parameter inversion problem, we chose
the fastest algorithm, SLSQP, which is second in accuracy.

\subsection{Fuzzy clustering}

For ionogram clustering, a variant of the well-known Expectation-Maximization(EM)
algorithm \cite{Dempster_1977} is used, with the track
model as a kind of mean value for the ionogram points distribution. Its
main difference from the standard EM algorithm, used, for example,
for classification using a mixture of normal distributions discussed above, is 
more complex dependence of the distribution shape on its hyperparameters:
the model distributions are constructed not for points on a plane,
but for their non-negative distances from certain parametric curves
on the plane. Thus, the desired model distribution of points is specified
for the coordinates of these points not directly, but parametrically.
Due to the representing the distributions
in new coordinates (distances from parametric curves) and the nonlinearity
of this transformation, the resulting clusters appear as curved stripes
on a plane and can intersect or contact each other, which allows to
solve the problem of separating ionogram tracks effectively.

The model distribution of points relative to each track is constructed
as follows. A significant parameter used to determine the probability
of a point belonging to this distribution is its deviation from the
model track shape (\ref{eq:track_model}). The distance is assumed
to be Euclidean, for height and frequency normalized by their 
standard deviations $\sigma_r,\sigma_f$ accordingly:

\begin{equation}
\begin{array}{c}
\Delta y_{i,m}=min_{j}\left\Vert \left(\tilde{f}_{i},\tilde{r}_{i}\right),\left(\tilde{f}_{j},\tilde{R}_{model}(f_{j}|\theta_{m})\right)\right\Vert _{2}\\
\Delta y_{i,m}\sim\mathcal{N}^{(+)}(0,\sigma_{m})\\
\tilde{f}_{i}=\frac{f_{i}}{\sigma_{f}};\tilde{r}_{i}=\frac{r_{i}}{\sigma_{r}};\tilde{R}_{model}=\frac{R_{model}}{\sigma_{r}}
\end{array}\label{eq:mean_em}
\end{equation}

Here we assume that $\Delta y_{i,m}\geq 0$ follows the half-normal
distribution $\mathcal{N}^{(+)}$ \cite{Daniel01111959}:

\begin{equation}
\mathcal{N}^{(+)}(x|0,\sigma)=\left\{ \begin{array}{l}
2\mathcal{N}(x|0,\sigma),x\geq0\\
0,x<0
\end{array}\right.
\label{eq:halfnorm_dist}
\end{equation}

To speed up calculations, the normalized
frequencies for calculating the distance $\Delta y$ are taken not
over the entire range of normalized frequencies, but relatively close
to $f_{i}$:

\begin{equation}
\left|\tilde{f}_{i}-\tilde{f}_{j}\right|<0.3
\label{eq:search_limits}
\end{equation}

Since the width of the range of variation of $\tilde{f}$ due to normalization
by the standard deviation is about 6, the use of the restriction (\ref{eq:search_limits}) 
allows
us to speed up the algorithm by approximately 5-10 times compared
to a complete search over all the points.

Experiments have shown that the use of euclidean distance from track
provides more robust clustering results than calculations based solely
on $r$ at a fixed frequency $f$, as used in track fitting (\ref{eq:track_model_fit_loss}),
although it is slower. It should be noted that the $\Delta y_{i,m}$
distance-finding algorithm is a very resource-intensive part of the
algorithm, and therefore a faster implementation is necessary. In
Python language used, the implementation was accelerated using the
numba library.

Reducing the frequency search range to 0.3 reduces computation time.
It should be noted that choosing a very small range can be reduced
to the classical least-squares method, for example (\ref{eq:SimpleFit}):

\begin{equation}
\begin{array}{c}
\Delta y(r_{i},f_{i}|\theta_{m})=\left|\tilde{r}_{i}-\tilde{R}_{model}(f_{i}|\theta_{m})\right|\\
\Delta y_{i,m}\sim\mathcal{N}^{(+)}(0,\sigma_{m})
\end{array}\label{eq:mean_em-1}
\end{equation}

is equivalent to:

\begin{equation}
\begin{array}{c}
\xi_{i,m}=\tilde{r}_{i}-\tilde{R}_{model}(f_{i}|\theta_{m})\\
\xi_{i,m}\sim\mathcal{N}(0,\sigma_{m})
\end{array}
\label{eq:mean_em-1-1}
\end{equation}

As numerical experiments have shown, the range (\ref{eq:mean_em-1-1}) should not be set
below 0.25-0.3 due to a significant fall of resulting clustering quality.

The weighted standard deviation of the points from the model can be
determined experimentally for all the ionogram points included into the
track, taking into account the half-normal distribution for $\Delta y_{i,m}$,
and taking into account their amplitude $A_{i}=A(f_{i},r_{i})$:

\begin{equation}
\sigma_{m}=\sqrt{\frac{\sum_{i=1}^{N_{m}}A_{i}\Delta y_{i,m}^{2}}{\sum_{i=1}^{N_{m}}A_{i}}}\label{eq:mse_em}
\end{equation}

At the expectation(E) stage, the expected probability that a point
belongs to a distribution calculated for a given track shape is
determined. Particular track for every point is choosen as the most
probable track with parameters $\theta_{m}$ and $\sigma_{m}$ defined
for each track $m$ at the modification stage.

At the modification(M) stage, new parameters $\theta_{m}$ are calculated
for the points belonging to a given track using the SLSQP method,
and the search algorithm described in the previous section. The differences
from standard algorithms are as follows.

The first difference is that the probability that point belongs to
a track is calculated by taking into account the points uniformity
($\rho$) on the track:

\begin{equation}
P(\Delta y)=\mathcal{N}^{(+)}\left(\Delta y\left(f,r|\theta_{m}\right)|0,\sigma\cdot(\rho+\varepsilon)\right)\label{eq:pdf_with_denisity}
\end{equation}

Here $\rho\in[0,1]$ is the average point uniformity on the track,
$\varepsilon=10^{-7}$ is a small bias for the stability of the numerical
algorithm. Experiments have shown that the use of point density allows
to increase the probability of dividing into tracks with more uniform
point distributions along the tracks in comparison with the tracks
with less uniform point distributions.

The average uniformity of points on a track $\rho$ is defined as
follows. The track in the interval between the minimum and maximum
frequency of points belonging to the track ($[min(f_{i}),max(f_{i})]$),
is equidistantly divided into $\sqrt{N}$ intervals, where N is the
number of points on the track. The uniformity $\rho$ is defined as
the relation of non-empty intervals number to total number of intervals. 
The dependence (\ref{eq:pdf_with_denisity})
is empirical and allows us to reduce the probability that a point
belongs to a track, due to an equivalent proportional decrease in
$\sigma$ and, accordingly, the probability in the case when the
points of the track are ununiformly distributed
along it. As shown experiments this allows us to improve the subjective
quality of track clustering.

The second difference is that the points belonging to each track at
this stage are determined by probabilistic Top2 sampling of the track
numbers (taking into account that, for physical reasons, typically
no more than two tracks intersect at a single point, and more complex
situations are rare). 
This means instead of choosing the most probable track for each point, we choose tracks
for each point randomly, based on some probability that depend on exact point and exact track.
The sampling probability is also determined
from the distribution for $\Delta y$ (\ref{eq:pdf_with_denisity}). Experiments
have shown that probabilistic sampling leads to more stable results
than selecting the most probable track for each point, since it allows
for multiple tracks to be traced over the same points, thereby more
reliably separating intersecting tracks.

The SLSQP algorithm, chosen for fitting the parameters of each individual
track over its points, is the fastest of the algorithms tested on
experimentally observed tracks, making it possible to use it within
the EM algorithm (which is also iterative). The standard deviation
$\sigma_{m}$ (\ref{eq:mse_em}) of the track points from the track
model is calculated over all the points that belongs the the track after the track parameters fit.

As experiments have shown, one of the ways to increase the stability
of the algorithm was a slight exponential smoothing of the parameters
$\theta_{m},\sigma_{m}$ at the end of each (i-th) iteration of the
EM algorithm:

\begin{equation}
\begin{array}{l}
\sigma_{m}^{(i+1)}\leftarrow(1-\beta)\sigma_{m}^{(i+1)}+\beta\sigma_{m}^{(i)}\\
\theta_{m}^{(i+1)}\leftarrow(1-\beta)\theta_{m}^{(i+1)}+\beta\theta_{m}^{(i)}
\end{array}\label{eq:sigma_median}
\end{equation}

where $\beta=0.1$ is choosen experimentally.

When initializing the algorithm based on a given initial clustering
of points, step M(modification) determines the initial values of the
parameters $\theta_{m}^{(0)},\sigma_{m}^{(0)}$ for each track. The
final combination of parameters $\theta_{m}^{(S)},\sigma_{m}^{(S)}$ is
found after S iterations of EM algorithm.

The resulting clustering model is a sum of $T$ distributions $P_{m}$
defined parametrically:

\begin{equation}
\begin{array}{l}
(f,r)\sim\sum_{m=1}^{T}\eta_{m}P_{m}\left(f,r\left|\theta_{m},\sigma_{m}\right.\right)\\
P_{m}\left(f,r\left|\theta_{m},\sigma_{m}\right.\right)=\mathcal{N}^{(+)}\left(\Delta y\left(f,r|\theta_{m}\right)\left|0,\sigma_{m}\cdot(\rho_{m}+\varepsilon)\right.\right)\\
\sum_{m=1}^{T}\eta_{m}=1
\end{array}\label{eq:em_model}
\end{equation}

where $\Delta y_{i,m}\geq 0$ is defined by the expression (\ref{eq:mean_em})
and have the half-normal distribution for the
normal distribution with zero mean and standard deviation $\sigma_{m}$.
The parameters $\eta_{m},\sigma_{m},\theta_{m}$ are determined by
fitting, the uniformity of points on the track $\rho_{m}$ is also
calculated inside the EM algorithm and is not a parameter. Thus, the
shape of T model distributions $P_{m\in[1,T]}$ is given parametrically
- not for the variables $f,r,A(f,r)$ themselves, but for their deviations
$\Delta y$ from T parametrically specified track models $R_{model}$
with parameters $\theta_{m}$.

As a result of the EM algorithm, for each identified track (cluster) $m$, we
obtain the probabilities of each ionogram point belonging to that
track, as well as the track parameters ($\theta_{m},\sigma_{m}$),
some of which are parabolic ionospheric layer parameters.
Thus, as a result, we not only cluster the data but also obtain some
equivalent ionospheric parameters for each detected track in a layer model
close to the parabolic one.

The EM algorithm is trained for no more than 150 iterations (epochs). The stopping
condition is determined by the minimum of the neg-log-likelihood
function and is detected by the increase in the neg-log-likelihood
function over 10 EM iterations.

Since each iteration of the EM algorithm uses an iterative SQLSP algorithm
to search for track parameters, this significantly slows down the
algorithm. However, as will be shown below, modern multithreaded personal
computers allow the problem to be solved in a time acceptable for
researchers.

\subsection{The optimal number of tracks}

A key problem in any clustering is the choice of its primary clustering
hyperparameter - the number of clusters. In most previous papers,
the number of clusters (tracks) was assumed to be fixed. In our approach,
we find the number of tracks, optimal for describing the ionogram,
from the ionogram itself using statistical critera.

As one of initial approximations for the EM algorithm, a separation
the ionogram points into weakly (or none) intersecting tracks looks usefull. 
For this purpose, when searching over T, one of the clustering variants 
is the model clustering using the superclustering method GMsDB \cite{Berngardt_2023},
which automatically determines the minimum possible number of weakly-intersected
clusters and can perform data labeling. The resulting number of clusters
$T_{g}$ and its labeling are used as the initial approximation of
the EM algorithm for $T=T_{g}$. The search of clusterings over T 
is carried out from T=2 to a sufficiently large number. 
To search for the optimal number of tracks $T_{opt}$, the number of
clusters $T_{i}$ increases by 1 each time and the initial labeling
is chosen randomly - with the numbers of point clusters uniformly
distributed in the range $[1,T_{i}]$. Except for the case $T_{i}=T_{g}$
- in this case, the initial labeling is chosen in accordance with
the clustering results by GMsDB algorithm \cite{Berngardt_2023}. 
As will be shown below,
the number of clusters $T_{i}$ is an upper limit on the number of
non-empty clusters $M_{i}$ found by the EM algorithm. The actual number $M_{i}$ 
may be smaller than $T_{i}$. By iterative increasing
$T_{i}$ and analyzing the clustering results, we can find the number
of clusters $T_{opt}$ that corresponds to the best clustering.

To determine $T_{opt}$, which provides the best clustering of the ionogram
into tracks, an internal clustering quality analysis is used. Since
our proposed method is model-based, an obvious and generally accepted
approach is to use one of the information criteria, such as the Bayesian
Information Criterion (BIC) or the Akaike Criterion (AIC). The Bayesian
Information Criterion and Akaike Criterion are easy to calculate for
this model, since the number of its parameters is known:

\begin{equation}
\begin{array}{c}
BIC(T)=2\cdot IC+\alpha T\left(Dim(\theta_{m})+1\right)ln(N)\\
AIC(T)=2\cdot IC+\alpha T\left(Dim(\theta_{m})+1\right)N\\
IC(T)=-ln\left(\sum_{m=1}^{T}\eta_{m}\mathcal{N}^{(+)}\left(\Delta y_{m}|0,\sigma_{m}\cdot(\rho_{m}+\varepsilon)\right)\right)
\end{array}\label{eq:bic_orig}
\end{equation}

Here $Dim(\theta_{m})=6$ is the number of parameters in the model
$R_{model}$, adding 1 corresponds to the model parameter $\sigma_{m}$.
In the classical version of BIC and AIC, $\alpha=1$. 

The second term in BIC and AIC (\ref{eq:bic_orig}) is a penalty for
model complexity. Since the fitting of the model with the initial
T models detects only M non-empty tracks ($M\leq T$), it was decided
to additionally take this into account when calculating the Bayesian
penalty by increasing $\alpha$ when $M\le T$:

\begin{equation}
\alpha=\frac{T}{M}\geq1\label{eq:apha_for_bic}
\end{equation}

\begin{figure}
\centering
\includegraphics[scale=0.5]{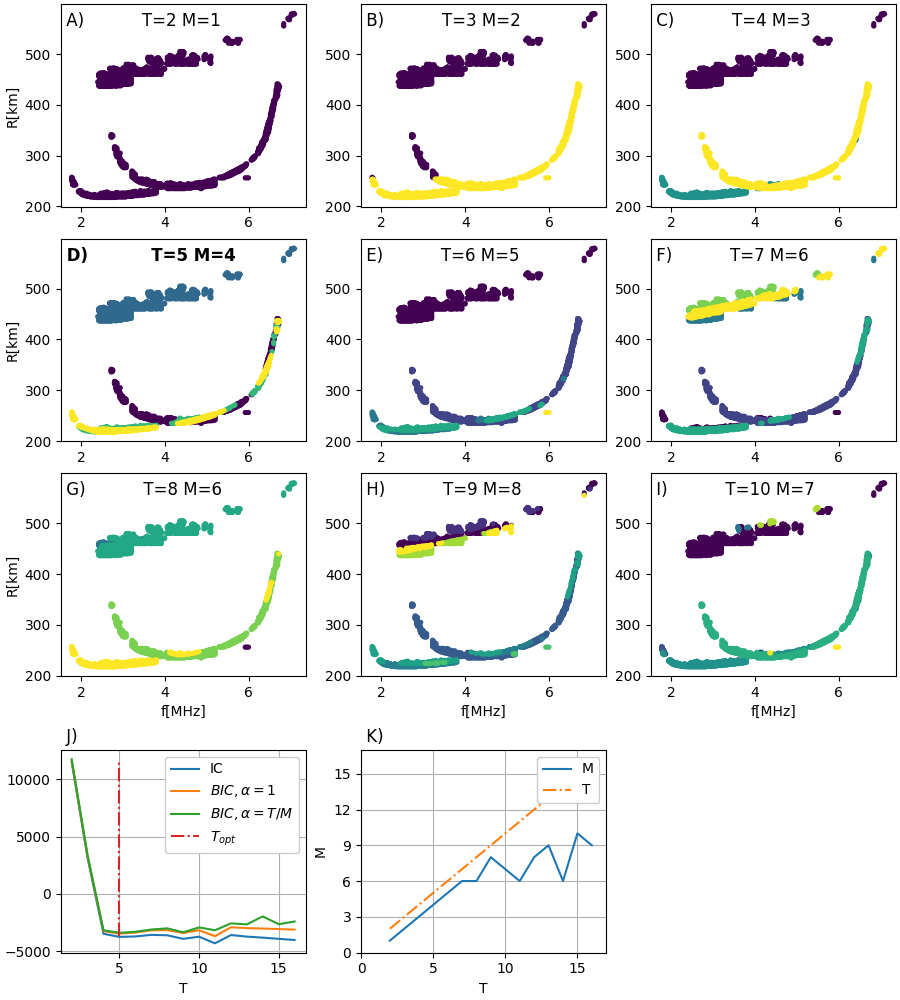}
\caption{An example of searching for optimal clustering for a filtered ionogram
without a part of the X-component. A-I) clustering for different numbers
of starting clusters T and the final number of non-zero clusters M,
optimal clustering is (D); J) dependence of the quality metrics IC
and BIC on T; K) dependence of the number of clusters M with a non-zero
number of points on the number of starting clusters T. The vertical
dash-dotted line in (J) is the position of the BIC minimum with $\alpha$
from (\ref{eq:apha_for_bic}).}
\label{fig:search_opt_T} 
\end{figure}

This empirical criterion additionally heavily penalizes the BIC and
AIC metrics for clusters that turn out to be empty as a result of
clustering, thereby preventing the model from making too many empty clusters.
An example of searching for optimal clustering is shown in Fig.\ref{fig:search_opt_T}.

Estimating the optimum based on the minimum AIC is unsuitable: the
number of detected tracks is too small - 1. Minimal BIC values ($T_{opt}=5$)
are suitable for estimating the optimum $T_{opt}$. The strong dispersion
of metrics depending on T is most likely due to the insufficient accuracy
of the EM and SLSQP algorithms and, as a result, the failure to achieve
the required approximation quality. Nevertheless, the BIC minimum
is clearly detected, allowing the fitting and clustering results
to be used in practice. Therefore, to find the optimal number of tracks
$T_{opt}$, we use the search for the minimum of $BIC(T)$. The minimum
is detected when BIC subsequently increases over 10 consecutive T
while searching over $T\in[2,\infty]$.

Thus, to determine the optimal number of clusters $T_{opt}$, we iterate
through T from 2 to a sufficiently large number (up to 28 in the current
implementation of the algorithm). For each number of clusters T, we
initialize and fit the algorithm and determine the BIC metric. Clustering
with the minimum BIC parameter value according to the Early Stopping
criterion (described above) with patience 10 is considered the optimal
clustering into tracks (clusters). The optimal training result is a clustering
of the ionogram into $M\leq T_{opt}$ clusters.

\subsection{The main part of the algorithm}

The main part of the algorithm is shown as the quasi-code in Algorithms
\ref{alg:FindOptimal}-\ref{alg:UpdateParams}.

The Algorithm \ref{alg:FindOptimal} is the main program, the purpose
of which is to cluster the ionogram at different T and determine the
optimal $T_{opt}$ and the corresponding clustering.

The Algorithm \ref{alg:EMsearch} is an implementation of the EM algorithm,
to cluster the data into tracks with maximum likelihood.

The Algorithm \ref{alg:UpdateParams} is an implementation of the
model parameter update in the EM algorithm, 
to optimally fit the model track curve over given cluster points.

\begin{algorithm}
\caption{Finding optimal clusters}
\label{alg:FindOptimal}

\begin{algorithmic}\par \Procedure {OptimalClustering}{$X$}

\hspace*{\algorithmicindent} \textbf{Input}: $X[:,3]$ - triplets
$f_{i},r_{i},A(f_{i},r_{i})$

\State $X_{o}\gets RemoveXcomponent(X)$

\State $labels\gets SuperclusteringUsingGMsDB(X_{o})$

\For {$T\in[2..28]$}

\If {$T=max(labels)$}

\State $L\gets labels$

\Else

\State $L\gets randomInt([1,T],len(X_{o}))$

\EndIf

\State $Distributions\gets InitTDistributions(T)$

\State$L,\theta_{m\in[1,T]},\sigma_{m\in[1,T]},BIC\gets EMsearch(X_{o},Distributions,L)$
using Algorithm \ref{alg:EMsearch}

\If {$BIC\,is\,minimum\,with\,patience\,10$}

\State $break$

\EndIf

\EndFor

\Return $L,\theta_{m\in[1,T]},\sigma_{m\in[1,T]}$

\EndProcedure

\end{algorithmic}
\end{algorithm}

~

\begin{algorithm}
\caption{Expectaion-maximization search}
\label{alg:EMsearch}

\begin{algorithmic}\par \Procedure {EMsearch}{$X,Distributions,Labels$}

\hspace*{\algorithmicindent} \textbf{Input}: $X[:,3]$ - triplets
$f_{i},r_{i},A(f_{i},r_{i})$

\hspace*{\algorithmicindent} \textbf{Input}: $Distributions$ - T
parametric distributions with parameters $\theta_{m},\sigma_{m}$

\hspace*{\algorithmicindent} \textbf{Input}: $Labels$ - labels for
X

\State $T\gets len(Distributions)$

\For {$m\in[1..T]$}

\State $W_{m}\gets len(X[Labels=m])/len(X)$

\EndFor

\For {$itteration\in[0..150]$}

\State \# E-step

\For {$m\in[1..T]$}

\State $Resp_{m,j}\gets GetP(X_{o,j},\theta_{m},\sigma_{m},\rho_{m})*W_{m}$,
using P from eq.(\ref{eq:pdf_with_denisity})

\EndFor

\State $Resp_{m,j}\gets Resp_{m,j}/\sum_{k}Resp_{k,j}$

\State \# M-step

\State $W_{m}\gets\sum_{j}Resp_{m,j}/len(Resp_{m,j})$

\State $\beta\gets0.1$

\For {$m\in[1..T]$}

\State $\theta'_{m},\sigma'_{m}\gets\theta{}_{m},\sigma{}_{m}$

\State $\theta_{m},\sigma_{m}\gets UpdateParams(X,Resp,m)$ using
Algorithm \ref{alg:UpdateParams}

\State $\theta_{m}\gets(1-\beta)\cdot\theta{}_{m}+\beta\cdot\theta'_{m}$

\State $\sigma_{m}\gets(1-\beta)\cdot\sigma_{m}+\beta\cdot\sigma'_{m}$

\State $\rho_{m}\gets GetTrackDensity(X,Resp_{m})$

\EndFor

\State $Labels\gets Argmax_{m}(Resp_{m,j})$

\State $M\gets len(Unique(L))$

\State $\alpha=T/M$, following eq.(\ref{eq:apha_for_bic})

\State $LogLikelihood\gets\sum_{j}ln\left(\sum_{m}W_{m}\cdot GetP(X,\theta_{m},\sigma_{m},\rho_{m})\right)$,
using P from eq.(\ref{eq:pdf_with_denisity})

\State $BIC\gets-2\cdot LogLikelihood+\alpha\cdot T\cdot(6+1)\cdot ln(len(X))$,
following eq.(\ref{eq:bic_orig})

\If {$LogLikelihood\,is\,maximum\,with\,patience\,10$}

\State $break$

\EndIf

\EndFor

\Return $Labels,\theta_{m\in[1,T]},\sigma_{m\in[1,T]},BIC$

\EndProcedure

\end{algorithmic}
\end{algorithm}

~

\begin{algorithm}
\caption{Updating distribution parameters}
\label{alg:UpdateParams}

\begin{algorithmic}\par \Procedure {UpdateParams}{$X,Resp,m$}

\hspace*{\algorithmicindent} \textbf{Input}: $X[:,3]$ - triplets
$f_{i},r_{i},A(f_{i},r_{i})$

\hspace*{\algorithmicindent} \textbf{Input}: $Resp_{m,j}$ - probabilities
for cluster m and samples $j$

\State $X_{s},Clusters\gets GetRandomSamplesForClusters(from\gets X,probability\gets Resp,mode\gets'Top2')$

\State $X_{s}\gets X_{s}[Clusters=m]$

\State $R,\theta_{m}\gets CirveFit(X_{s},method='SLSQP')$, using
eqs.(\ref{eq:track_model},\ref{eq:track_model_fit_loss}) and parameter
constrants.

\State $\Delta y_{m}\gets MinDistances(X_{s},R))$, using eq.(\ref{eq:mean_em})

\State $\sigma_{m}\gets GetSigma(\Delta y_{m},X_{s}[:,2]))$, using
eq.(\ref{eq:mse_em})

\Return $\theta_{m},\sigma_{m}$

\EndProcedure

\end{algorithmic}
\end{algorithm}

\section{Discussion}

\subsection{Advantages and disadvantages of the method}

The advantages of this method include its high adaptability, which
is particularly useful in disturbed conditions where
 predefined ionogram model with a fixed number
of tracks or parameters fails:
\begin{itemize}
\item When cleaning the ionogram from the noise, it automatically adapts
to the noise level;
\item When clustering ionogram into tracks, it can identify and combine
discontinuous tracks into a single one;
\item The method can separate intersecting or contacting tracks;
\item The method automatically adapts to an unknown number of tracks
on the ionogram.
\end{itemize}
Currently, the applicability of this method is limited to conditions
where the track shape can be described by the model (\ref{eq:track_model}),
and the ionosphere is correspondingly described by a model of several
well-separated parabolic layers, possibly semi-transparent. For more accurate track
separation and interpretation, more complex track models $R_{model}(f|\theta)$
should be used.

\begin{figure}
\centering
\includegraphics[scale=0.5]{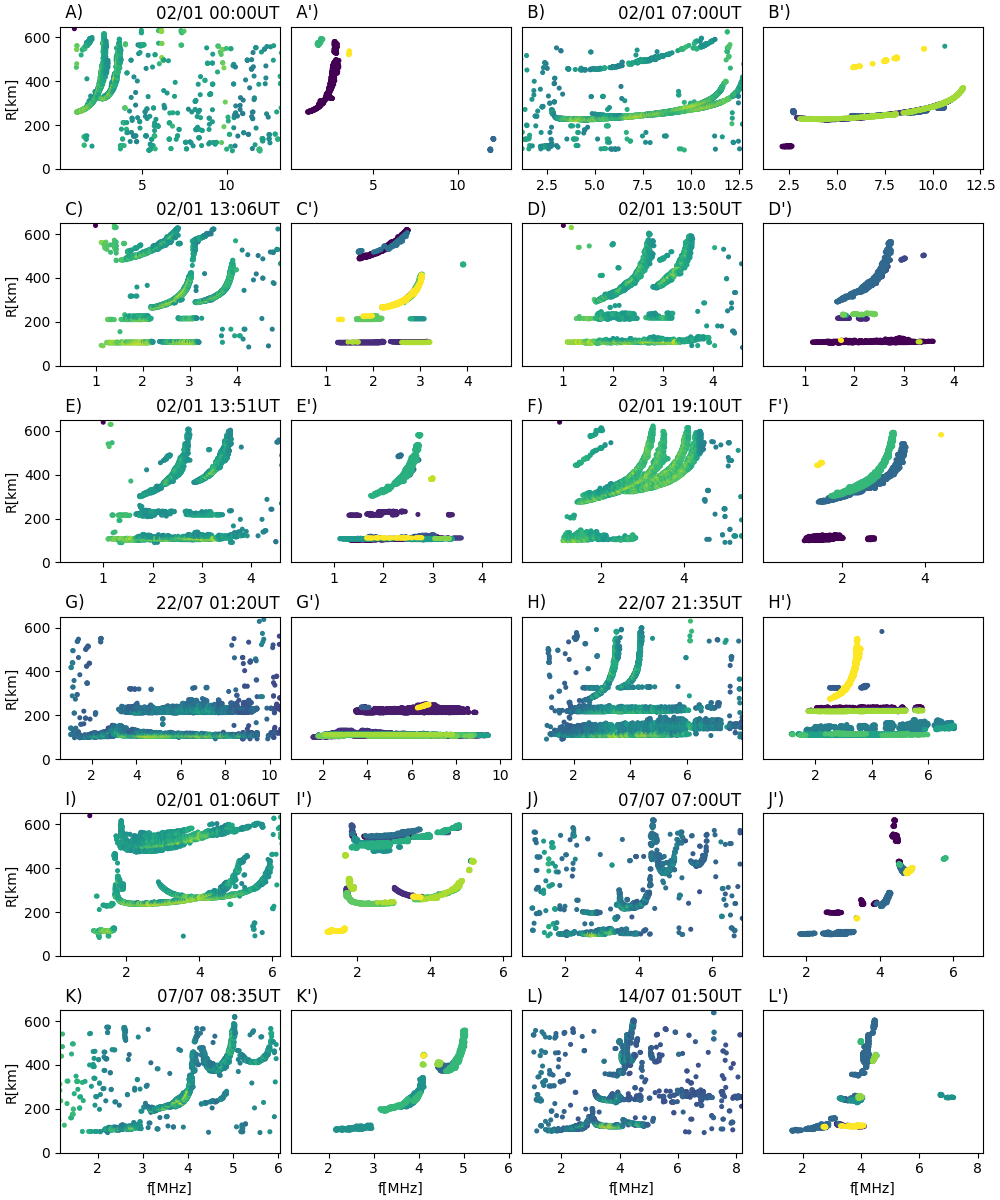}
\caption{The examples of clustered ionogrames. A-L) source ionogrames, 
the color corresponds to the signal intensity, A'-L') clustered ionogrames with X-component partly removed,
different colors corresponds to different clusters detected.}
\label{fig:cluster_examples}
\end{figure}

Examples of the clusterer's operation in various cases are shown in
Fig.\ref{fig:cluster_examples}. It is evident that the method is
quite efficient in the case of several semi-transparent layers, multiple
reflections, or horizontal inhomogeneities (Fig.\ref{fig:cluster_examples}A'-F').
The problem with the method is the extraction of tracks on highly
noisy ionograms with a small number of informative points (Fig.\ref{fig:cluster_examples}J',L'),
as well as complex cases of a multi-layer ionosphere or a large number
of complex tracks (Fig.\ref{fig:cluster_examples}I',K'). In this
case, the method tends to combine tracks from the E, F1 and F2 layers
into a single track instead of separating them (Fig.\ref{fig:cluster_examples}J',K',L').

A disadvantage of the method is the need to process only the O-component:
as analysis has shown, clustering both the O- and X-components within
a single 12-parameter model yields less stable results, especially
in the presence of underlying layers. Therefore, it is recommended to
limit the data by the O-component only and pre-remove a significant
portion of the X-component points before using the model. 
We use the following algoritm to detect O-mode points:

\begin{equation}
(f,r) \in O: 
\left\{
\begin{split} 
& \exists (f',r'): ( (f',r') \in [(f+f_h,r-10km); (f+f_h+0.25MHz),(r+10km)] ) \\
& or \\
& \nexists (f',r'): ( (f',r') \in [(f-f_h-0.25MHz,r-40km); (f-f_h,r)] ) \\
\end{split} 
\right\}
\label{eq:OXfilter}
\end{equation}
where $f_h$ = 0.75MHz corresponds to hyrofrequency mode separation.
The points that does not fit (\ref{eq:OXfilter}) are supposed to be X-mode ones.

This leads to the appearance of artifacts - clustering of undeleted portions
of the X-components - and the resulting errors. The model was tested
on ionosonde data without hardware separation of the O- and X-components;
on ionosondes with hardware separation of the components (for example,
by signal polarization \cite{Reinisch_2008_digisonde,Harris_2017}),
this problem may be less significant.

Another disadvantage of the method is its stochastic and iterative
nature - with the same input data, different results can be obtained
due to the stochastic nature of the EM algorithm used.

\begin{figure}
\centering
\includegraphics[scale=0.5]{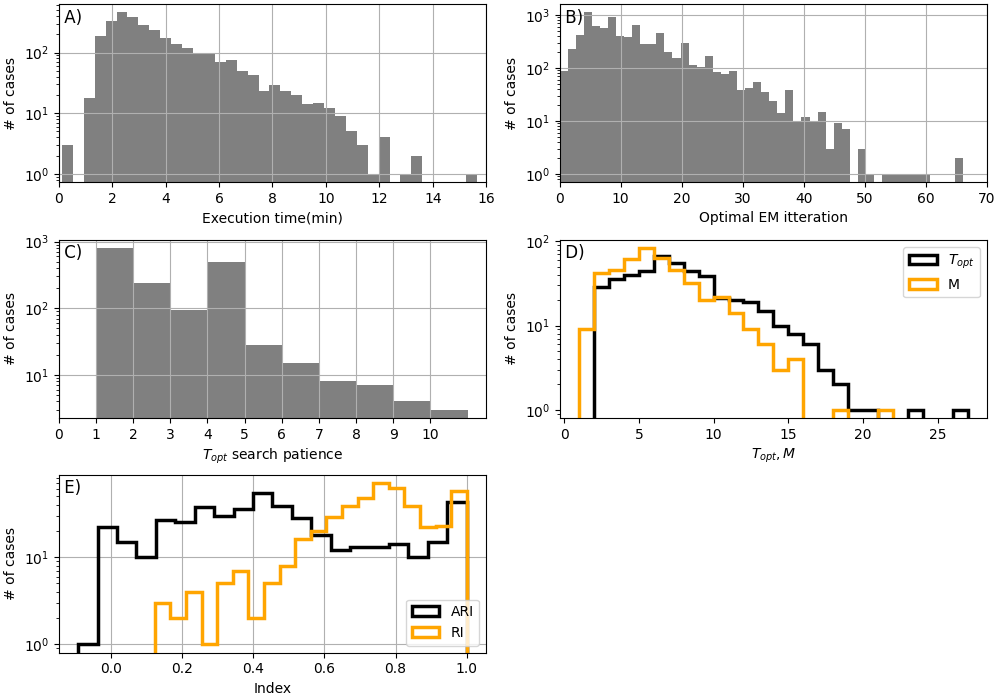}
\caption{A) Distribution of algorithm execution times over different ionograms;
B) Number of EM algorithm iterations over different ionograms; C)
Distribution of patience for searching $T_{opt}$ over different ionograms;
D) Distribution of optimal $T_{opt}$ and found non-zero classes M
over different ionograms; E) Comparison of the resulting clustering
with the result of GMsDB clustering (degree of isolation of ionogram
tracks) over different ionograms.}
\label{fig:time_efficiency} 
\end{figure}

The disadvantage of the method is its speed - the average time of processing
one ionogram with a 32-core processor is 3.7 minutes (Fig. \ref{fig:time_efficiency}A).
From Fig.\ref{fig:time_efficiency}B it is evident that, according
to the experimental data, the number of EM algorithm iterations usually
does not exceed 66. From Fig.\ref{fig:time_efficiency}D it is evident
that the optimal initial number of modes $T_{opt}$ does not exceed
27, and the final number of modes M does not exceed 22. This allows
us to limit search over T. The majority of cases can be fited with patience\textless 10
(Fig.\ref{fig:time_efficiency}C). Thus, to optimize the computing
time, it makes sense to limit the number of EM algorithm iterations
to approximately 55, and the maximum number of modes T to 18. Currently,
the algorithm is implemented in Python; it is possible that the use
of more efficient programming languages will also speed up the program.

A drawback of this method is the linear proportionality of the execution
time to the number of clusters T (which can be improved by using multithreaded
processors with high number of threads, so that all
clusters can be efficiently analyzed in parallel). Using brute-force
search for $T_{opt}$ leads to a quadratic dependence of the execution
time of the whole algorithm on the maximal number of clusters. One
could try replacing the brute-force search for T with faster search methods,
but the BIC quality metric has a large variability over T, and
therefore using faster methods based on the smoothness of this curve
looks not useful, and a exhaustive search looks preferable.

One of the issues is to understand if it is possible to 
interpret the tracks without using a special physically
based models. Obviously, this is easy to do if all the tracks on the
ionogram are separated from each other (Fig.\ref{fig:1}G,H). 
In this case, simple histogram separation methods,
such as \cite{LYNN2018} or clustering as \cite{Berngardt_2023},
can be used to separate tracks. This case corresponds to the best
of agreement between the clustering obtained without a physical track
model (in our case, we use GMsDB algorithm \cite{Berngardt_2023})
and the clustering obtained using our physically-informed algorithm.
The degree of similarity of the clusterings can be estimated
using the Adjusted Rand Index (ARI) metric, that reaches its maximum
1 when clusterings are identical. The results of such an analysis over 
a number of experiments are shown in Fig.\ref{fig:time_efficiency}E.
From the figure it is clear that in most cases
the clustering obtained using our model method differs from the non-intersecting
track model, so in the most cases the clustering not based on physical models will
fail.

\subsection{Qualitative analysis of clustering results}

\begin{figure}
\centering
\includegraphics[scale=0.5]{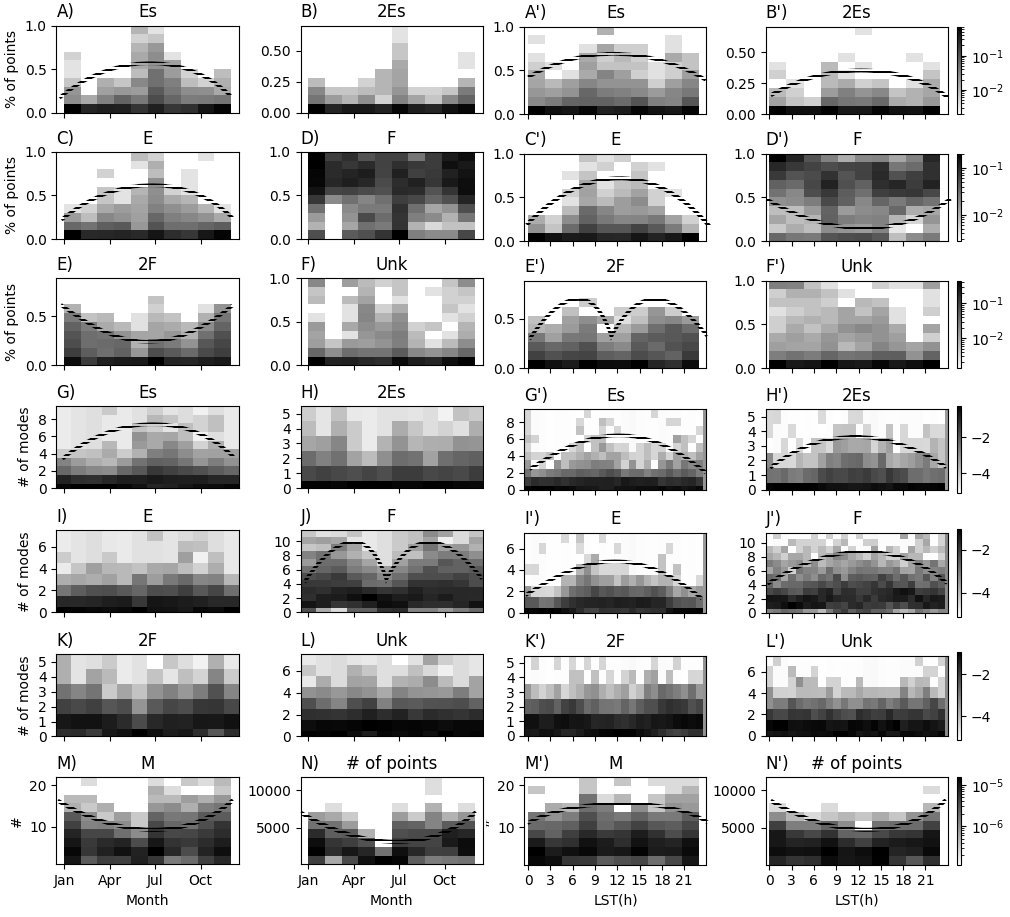}
\caption{Estimation of relative quantity and types of different layers on the
ionogram, as well as the number of modes and the number of informative
points depending on the month of the year (A-N) and on the local solar
time (A'-N'). A,G,A',G') - Sporadic E; B,H,B',H') - Sporadic E of
the second hop; C,I,C',I') - E-layer; D,J,D',J') - F-layer; E,K,E',K')
- F-layer of the second hop; F,L,F',L') - uninterpreted data; M,M')
- the number of detected modes; N,N') - the number of informative
points. The dotted line, where possible, highlights the estimate of
the trends in parameter changes depending on the season or local solar
time.}
\label{fig:mode_stats}
\end{figure}

A useful application of this method is the ability to estimate the
quantity and quality of various layers based on clustering results.
For example, a large number of sporadic E or F multiple reflections,
maximum frequencies of sporadic layers, or their wide spread. With
this method, tracks from different layers are often already separated,
so the characteristics of each cluster can be used for its preliminary
interpretation, similar to histograms \cite{LYNN2018}. For example,
cluster thickness can be used to search for sporadic E layers - they
are usually thin, but can be located at different distances depending
on the reflection hop (which can be estimated from their range).
The combination of minimum height and cluster thickness can be used
to search for F layers of various hops - the thickness of these layers
is often significant, over 50 km, and minimum heights over 400 km
correspond to second-hop scattering. The presence of multiple clusters,
defined as the F layer, but with different maximum frequencies may
indicate the presence of horizontal inhomogeneities. The advantage
of this method over histograms \cite{LYNN2018} is that after clustering,
each track is already detected and isolated, and separating intersecting
components in histograms is not necessary. In simple
cases (tracks are isolated by height), these methods should produce
similar results. The results of statistical processing of annual observations
at the Tory observatory by our method and preliminary intepretation,
described above, are shown
in Fig.\ref{fig:mode_stats}.

From Fig.\ref{fig:mode_stats}D it is evident that the F layer of
the first hop is observed almost constantly on ionograms, with a decrease
in the summer months associated with increasing absorption in the
E-layer and sporadic E-layer. E and Es laysers during this period
demonstrate a significant increase (Fig.\ref{fig:mode_stats}A,C).
The number of informative points (Fig.\ref{fig:mode_stats}N) demonstrates
decrease during the absorption periods. An increase in
the number of Es of the second hop is observed in the winter months and July.

The number of modes demonstrates a rather insignificant seasonal dependence (Fig.\ref{fig:mode_stats}G-L),
increasing during daylight hours (Fig. \ref{fig:mode_stats}M'), as
does the number of single (Fig.\ref{fig:mode_stats}A') and multiple
reflections from Es (Fig. \ref{fig:mode_stats}B'), reflections from
E (Fig. \ref{fig:mode_stats}C'), and multiple reflections from F (Fig.\ref{fig:mode_stats}E').

The shown results could be sometimes incorrect - sometimes we can
incorrectly exclude X-component from the data, and without accurate
processing the resulting tracks with more accurate inversion models,
taking into account all the found layers, it is difficult to interpret
them. In combination with previously developed profile reconstruction
algorithms \cite{Ponomarchuk2026} in a simple model ionosphere, the clustering method can
enable the processing of profiles not only in a quiet ionosphere but
also in a disturbed ionosphere, including the determination of parameters
that do not fit into simple models - the number and heights of
sporadic layers, including semi-transparent ones, the presence of
horizontal disturbances and their characteristic scale, and the detection
and interpretation of multiple reflections.

\section{Conclusion}

This paper presents a physically-informed fuzzy clustering of tracks
on vertical sounding ionograms and determining their
optimal number for automatically separating the ionogram
into tracks suitable for further interpretation. 
The model is developed for use not only in ionospheric
conditions where the number of tracks is known, but also in disturbed
conditions where the number of tracks is initially unknown.

The proposed method (Algorithms \ref{alg:FindOptimal}-\ref{alg:UpdateParams})
is based on the EM algorithm, which enables fuzzy model clustering
and on parametrically defined distributions of distances from points
to parametrically specified curves. The track model is based on curves
similar to the model tracks in the parabolic ionospheric
layer model, additinaly complicated (\ref{eq:track_model}) to enable
separate analysis of the contribution of each layer in the presence
of underlying layers. The final model of each track has six parameters:
three standard ones (the critical frequency, the lower boundary of
the layer, and its half-width in the parabolic layer model), and three
additional ones to account for the influence of possible underlying
layers on the studied track. By iterating through the number
of tracks, the model finds the optimal number
of tracks on the ionogram, based on the minimum of modified Bayesian
information criterion. The SLSQP algorithm
is used to find track parameters, as it is the fastest and acceptably
accurate algorithm for this task.

To improve the quality of ionogram clustering, automatic adaptive
noise filtering is performed before clustering. This filtering is
based on a combination of the DBSCAN and Gaussian Mixture algorithms.
The former is used to identify noise-like outliers, and the latter
is used to find the first algorithm's hyperparameter $\varepsilon_{opt}$
(\ref{eq:eps_prefilter}). Also, to improve clustering quality, a
rough preliminary removal of points belonging to the extraordinary
radiowave is made.

The algorithm allows separating contacting and intersecting tracks,
as well as merging discontinuous tracks into one. The average performance
of the Python implementation on a 32-thread processor with 5 GHz clock
frequency is 3.7 minutes per ionogram, making it suitable for real-time
use with appropriately frequent ionogram acquisition.

A disadvantage of the method is the need to process only the O-component.
On ionosondes with hardware-based component separation \cite{Reinisch_2008_digisonde,Harris_2017},
this problem may be less significant. Another disadvantage of the
method is the linear proportionality of the execution time to the
number of clusters. When using an exhaustive search for $T_{opt}$,
this leads to a quadratic dependence of the execution time on the
maximum number of clusters. Another disadvantage of the method is
its stochastic nature.

The advantages of this method include its high degree of adaptability,
which is particularly useful in disturbed conditions, where standard
interpretation methods based on a predefined ionogram model with a
fixed number of tracks or parameters fail. Currently, the applicability
of this method is limited to conditions where the track shape can
be described by the model (\ref{eq:track_model}), and the ionosphere
is correspondingly described by a model of several well-separated semi-transparent
parabolic layers. For more accurate track separation and interpretation, 
the more complex track models $R_{model}(f|\theta)$ should be used.

Statistical processing of annual observations allows us to identify
explainable seasonal and daily dynamics of observation of various
layers (F, E, Es) and their multiple reflections (Fig.\ref{fig:mode_stats}),
without the use of subsequent physical inversion of tracks by other
algorithms.

\section*{Acknowledgments}

The results were obtained using the Core Shared Research Facility "Angara" 
of ISTP SB RAS (https://ckp-rf.ru/catalog/ckp/3056/). 
The work was financially supported by the Ministry of Science and Higher Education.



\end{document}